\documentclass[showpacs,preprintnumbers,amsmath,amssymb]{revtex4}

\usepackage{graphicx}
\usepackage{subfigure}
\usepackage{dcolumn}
\usepackage{bm}

\begin{document}

\title{Microscopic expression of the second law of thermodynamics}
\author{Takaaki Monnai}%
\email{monnai@suou.waseda.jp}%
\affiliation{$*$Department of Applied Physics, Osaka City University,
3-3-138 Sugimoto, Sumiyoshi-ku, Osaka 558-8585, Japan}
\begin{abstract}
We give an explanation of the second law for macroscopic quantum systems under the assumption that the initial state is described by a canonical ensemble.
We formally derive the positivity of the nonequilibrium entropy difference.
Then it is shown that the nonequilibrium entropy difference can be regarded as the entropy difference in thermodynamics for typical relaxation processes of macroscopic quantum systems.        
\end{abstract}
\maketitle
\section{Introduction}
Since the pioneering work by L.Boltzmann, considerable attentions have been 
paid to the microscopic derivation of the second law of thermodynamics as shown in many literatures such as Ref.\cite{Zwanzig}.
These seminal works explain the relaxation to the equilibrium as a tendency toward ``the most probable state".
The relaxation processes are irreversible as expressed by the so-called H-theorem based on the kinetic entropy. 
However, the microscopic time reversibility and Poincar\'e recurrences\cite{Zwanzig} raised crucial criticisms to such probabilistic interpretations, i.e., the entropy can not monotonically increase in the course of time evolution.
In order to avoid recurrent behaviors, infinitely large reservoirs are often employed\cite{Karpov,Tasaki} so that the recurrence time goes to infinity.      
However the assumption of infinite reservoirs would be unrealistic and excludes framework based on the partition function\cite{Tasaki}.
On the other hand, it is remarkable that even for finite quantum systems, there is a well-defined non-negative entropy production which expresses a correlation between the subsystem and finite reservoirs as a relative entropy\cite{Esposito1}.

In this article, we straightforwardly derive a microscopic expression of the second law of thermodynamics for macroscopic quantum systems by taking into account for the relaxation to a canonical state in the evaluation of expectation values\cite{Monnai}.
By assuming the phenomenological law of relaxation property which is verified for the expectation value of macroscopic quantities, we reveal {\it the microscopic expression of the entropy increase}. 
It is shown that the entropy difference between the initial and final states is expressed as a  relative entropy which measures a lag between the actual final state and approximate canonical state. 
\section{Second law of thermodynamics}
Let us consider a macroscopically large but finite initially isolated system. The initially isolated system is typically composed of a subsystem and a reservoir. 
After initial time $t=0$, there is an external forcing acting on the system, and the total Hamiltonian $H(t)$ depends on time.  
Until $t=0$, the density matrix describing the initial state is assumed to be canonical
\begin{equation}
\rho(0)=\frac{1}{Z(0)}e^{-\beta H(0)},\label{canonical}
\end{equation}
where $\beta=\frac{1}{k_B {\rm T}}$ is the inverse of the temperature ${\rm T}$, $k_B$ is the Boltzmann constant, 
 and $Z(0)$ is the partition function determined from the normalization, respectively.
In order to avoid confusion, it is pointed out that the canonical ensemble can describe the isolated systems as well as open systems, i.e. the density of energies has an extremely sharp peak and the canonical ensemble is practically yields the same value as the corresponding microcanonical ensemble.
The use of canonical ensemble is reasonable especially for the case that the total system is composed of a subsystem and a large reservoir.
In the course of time evolution, the external work is done on the system through the time dependence of $H(t)$.

The deterministic external forcing acts during the time interval $0\leq t \leq T_0$, and switched off for $t>T_0$. 
It is expected that after a sufficiently long waiting time, i.e. at $t={\it T}\gg T_0$, the density matrix reaches a state which yields approximately same expectation values as the canonical ensemble $\rho({\it T})=\frac{1}{Z({\it T})}e^{-\hat{\beta} H({\it T})}$ at an inverse temperature $\hat{\beta}=\frac{1}{k_B\hat{{\rm T}}}$.
Note that even when the actual state does not relax to equilibrium, $\rho({\it T})$ and corresponding partition function are well-defined. 
Further microscopic justification of the assumption is described below Eq.(8). 

Then we calculate the entropy difference $\Delta S$ between the initial and final equilibrium states.
Here according to the statistical mechanics, the entropy is regarded as the logarithm of the number of the states $\Omega(E(s))\delta E$ which have a constant energy $E(s) \;(s=0,T)$ with a small uncertainity $\delta E\ll E(s)$. 
First, we give an expression of the entropy difference. 
As the equilibrium statistical mechanics, let us define the entropy at $t=T$ as 
\begin{eqnarray}
\frac{S({\it T})}{k_B}&\equiv& {\rm Tr}\{U\rho(0)U^+\hat{\beta}H({\it T})+\log Z({\it T})\} \nonumber \\
&=&\hat{\beta}(\langle H({\it T})\rangle-F(T)), \label{free}
\end{eqnarray}
where $F(T)=-\frac{1}{\hat{\beta}}\log Z({\it T})$ is the free energy at time $T$. $F(T)$ is specified by the temperature $\hat{\beta}$ and Hamiltonian $H(T)$.  
\begin{equation}
U={\rm T}\{e^{-\frac{i}{\hbar}\int_0^T dt H(t)}\}
\end{equation}
 is the unitary time evolution operator expressed by the time-ordered product.
Therefore, we can rewrite the entropy difference 
\begin{equation}
\Delta S\equiv S({\it T})-S(0)
\end{equation}
 as  a relative entropy between the actual final state $U\rho(0)U^+$ and the canonical state $\rho({\it T})$
\begin{eqnarray}
&&\frac{\Delta S}{k_B} \nonumber \\
&=&\left({\rm Tr}\{U\rho(0)U^+\hat{\beta}H({\it T})\}-\hat{\beta}F({\it T})\right)-\left({\rm Tr}\{\rho(0)\beta H(0)\}-\beta F(0)\right) \nonumber \\
&=&-{\rm Tr}\{U\rho(0)U^+\left(\log\rho({\it T})-\log U \rho(0)U^+\right)\} \nonumber \\
&=&D[U\rho(0)U^+||\rho({\it T})] \nonumber \\
&\geq& 0, \label{entropy}
\end{eqnarray}
which is by definition non-negative\cite{Thirring,Esposito1}.
Note that $\Delta S$ is a thermodynamic entropy difference only when the state $U\rho(0)U^+$ is an equilibrium state.
Remarkably, the entropy change describes how actual and equilibrium states are different.  
This is one of the results of this article. 

Below we evaluate the typical value of $\Delta S$.     
As shown in Ref.\cite{Monnai}, we can derive a thermalization of the expectation value of an observable $A$ which polynomially depends on the system size
\begin{equation}
{\rm Tr}U\rho(0)U^+A\cong{\rm Tr}\rho({\it T})A. \label{relaxation}
\end{equation}
The actual state $U\rho(0)U^+$ and the equilibrium state $\rho({\it T})$ are close in such a way.
In order to estimate the accuracy of the approximation, we outline the derivation\cite{Monnai}.  
The actual final state is reached by a unitary time evolution
\begin{eqnarray}
&&U\rho(0)U^+ \nonumber \\
&=&e^{-\frac{i}{\hbar}H({\it T})(T-T_0)}\rho(T_0)e^{\frac{i}{\hbar}H({\it T})(T-T_0)} \nonumber \\
&=&\sum_{n,m} c_{n,m} |E_n\rangle\langle E_m|, \label{evolution}
\end{eqnarray}
where $c_{n,m}=e^{-\frac{i}{\hbar}(E_n-E_m) (T-T_0)}\langle E_n|U(T_0)\rho(0)U(T_0)^+|E_m\rangle$, and $U(T_0)=T\{e^{-\int_0^{T_0}\frac{i}{\hbar}H(t)dt}\}$.
After $T_0$, the total system is isolated, and evolves by the Hamiltonian $H(T)$.  
On the other hand, the off-diagonal elements of physical quantity $A$ is negligible compared to the diagonal elements:
\begin{itemize}
\item[i)] The off-diagonal matrix element $\langle E_n|A|E_m\rangle$ is typically negligible in the macroscopic limit.
Physically, this would mean that transition amplitudes between macroscopically different states due to the perturbation $A$ are extremely small.
We shall give a quantitative explanation of this statement. 
Let us diagonalize the quantity $A$ as 
\begin{equation}
A=\sum_n A_n |A_n\rangle\langle A_n|,
\end{equation}
and define its square root 
\begin{equation}
B=\sum_n \sqrt{A_n} |A_n\rangle\langle A_n|.
\end{equation}
We also define state vectors $\{|\Phi_n\rangle=B|E_n\rangle\}$ so that the matrix element is expressed as the inner product
\begin{equation}
\langle E_n|A|E_m\rangle=\langle\Phi_n|\Phi_m\rangle. \label{offdiagonal}
\end{equation}
The states $\{|\Phi_n\rangle\}$ are chosen from the extremely large Hilbert space ${\cal H}$ with various directions.  
Thus we assume that the sequence of the normalized vectors $\{\frac{|\Phi_1\rangle}{\sqrt{\langle\Phi_1|\Phi_1\rangle}},\frac{|\Phi_2\rangle}{\sqrt{\langle\Phi_2|\Phi_2\rangle}} ,...\}$ is regarded as a {\it uniformly random sampling} from ${\rm dim}{\cal H}$ dimensional unit sphere.    
It is then straightforward to show that the mean square of the inner product is smaller than $\frac{\|A\|^2}{{\rm dim}{\cal H}}$\cite{Lebowitz}, which we will show later.
Here $\|A\|$ is the maximum of the eigenvalues of $A$. 

Let us derive the inequality for the inner product
\begin{equation}
\langle|\langle\Phi_n|\Phi_m\rangle|^2\rangle\leq\frac{\|A\|^2}{{\rm dim}{\cal H}}, \label{inner}
\end{equation}  
where the bracket shows the average with respect to the uniform random sampling of $\frac{|\Phi_n\rangle}{\sqrt{\langle\Phi_n|\Phi_n\rangle}}$ from the unit sphere.
In Eq.(\ref{inner}), $\|A\|$ polynomially depends on the system size.
On the other hand, the dimension ${\rm dim}{\cal H}$ grows exponentially as the system size increases.
Then inequality (\ref{inner}) shows that the off-diagonal elements Eq.(\ref{offdiagonal}) is extremely small for the macroscopic system.
In order to derive the inequality (\ref{inner}), we introduce the vector representation of $\frac{|\Phi_n\rangle}{\sqrt{\langle\Phi_n|\Phi_n\rangle}}$ in a orthogonal complete basis as 
$\vec{d_n}=(\cos\theta_1,\sin\theta_1\cos\theta_2,\sin\theta_1\sin\theta_2\cos\theta_3,...,\sin\theta_1\sin\theta_2\cdot\cdot\cdot\sin\theta_{d-1})$ with $d={\rm dim}{\cal H}$.
Without loss of generality, another vector can be fixed at $\vec{d_m}=(1,0,0,...,0)$.
The square inner product is then calculated as
\begin{eqnarray}
&&\frac{\langle|\langle\Phi_n|\Phi_m\rangle|^2\rangle}{\langle\Phi_n|\Phi_n\rangle\langle\Phi_m|\Phi_m\rangle} \nonumber \\
&=&\langle(\vec{d_n}\cdot\vec{d_m})^2\rangle=N\int_0^{\frac{\pi}{2}}\cdot\cdot\cdot\int_0^{\frac{\pi}{2}}\cos^2\theta_1\bigl|\frac{\partial(r\cos\theta_1,...,r\sin\theta_1\cdot\cdot\cdot\sin\theta_{d-1})}{\partial(\theta_1,...,\theta_{d-1},r)}\bigr|_{r=1}d\theta_1\cdot\cdot\cdot d\theta_{d-1} \nonumber \\
&=&N\int_0^{\frac{\pi}{2}}\cdot\cdot\cdot\int_0^{\frac{\pi}{2}}(1-\sin^2\theta_1)\sin^{d-2}\theta_1\sin^{d-3}\theta_2\cdot\cdot\cdot\sin\theta_{d-2}d\theta_1\cdot\cdot\cdot d\theta_{d-1} \nonumber \\
&=&1-\frac{(d-2)\Gamma(\frac{d-2}{2})}{\Gamma(\frac{d-1}{2})}\frac{\Gamma(\frac{d+1}{2})}{\Gamma(\frac{d}{2})d} \nonumber \\
&=&\frac{1}{d}, \label{inequality}
\end{eqnarray}
where $N=\prod_{n=1}^{d-2}\frac{n\Gamma(\frac{n}{2})}{\Gamma(\frac{n+1}{2})}$ is the normalization factor.
Since $|E_n\rangle$ and $|E_m\rangle$ are normalized, 
\begin{eqnarray}
&&\langle \Phi_n|\Phi_n\rangle\langle\Phi_m|\Phi_m\rangle \nonumber \\
&=&\langle E_n|A|E_n\rangle\langle E_m|A|E_m\rangle \nonumber \\
&\leq&\|A\|^2. \label{normalization}
\end{eqnarray}
The inequality (\ref{inner}) derives from Eqs.(\ref{inequality},\ref{normalization}).
From Eq.(\ref{inner}), the coefficient $c_{n,m}$ of Eq.(\ref{evolution}) is safely replaced by $\langle E_n|U\rho(0)U^+|E_m\rangle \delta_{n,m}$ in the evaluation of ${\rm Tr}U\rho(0)U^+A$.
More quantitatively, the contribution from the off-diagonal elements is evaluated as
\begin{eqnarray}
&&\sum_{n\neq m}^d\langle E_n|U\rho(0)U^+|E_m\rangle\langle E_m|A|E_n\rangle \nonumber \\
&\cong& \frac{\|\rho(0)\|}{\sqrt{d}}\frac{\|A\|}{\sqrt{d}}\sum_{n,m}^d e^{i\phi_{nm}}, \label{offdiagonal}
\end{eqnarray}
where $\phi_{nm}$ is the phase for the $nm$ matrix element. 
If the quantities $U\rho(0)U^+$ and $A$ are completely uncorrelated, then we have 
\begin{equation}
\sum_{n,m}^d e^{i\phi_{nm}}=O(d),
\end{equation} 
and Eq.(\ref{offdiagonal}) is $\|\rho(0)\|\|A\|$.
Since $\|\rho(0)\|\cong\frac{1}{d}$ from the normalization, and the off-diagonal contribution is indeed negligible.
\item[ii)] It is remarked that the canonical distribution given in Eq.(\ref{canonical}) is highly degenerated at an energy $E$ for macroscopic systems, and therefore 
practically equivalent to the microcanonical ensemble $\frac{\delta(E-H(0))}{\Omega(E(0))}$. 
Therefore we first concern with the microcanonical ensemble, i.e. we replace $\rho(0)$ with $\frac{\delta(E-H(0))}{\Omega(E(0))}$ as 
\begin{equation}
\rho(0)\cong\frac{\delta(E-H(0))}{\Omega(E(0))}.
\end{equation}
Then the diagonal element is expressed as 
\begin{eqnarray}
&&\langle E_n|U\rho(0)U^+|E_n\rangle \nonumber \\
&\cong&\frac{1}{\Omega(E(0))}\langle E_n|\delta(UH(0)U^+-E)|E_n\rangle \nonumber \\
&=&\frac{1}{\Omega(E(0))}\sum_m \delta(E_m(0)-E)|\langle E_n|U|E_m(0)\rangle|^2, \label{element}
\end{eqnarray}
where we introduced the normalized eigenstates of $H(0)$ as $H(0)|E_n\rangle=E_n|E_n\rangle$.
In the relaxation process, $|\langle E_n|U|E_m(0)\rangle|^2$ would be non negligible only when the conservation of the energy is well-satisfied after the long waiting time, 
i.e. $E_n\cong E_m(0)+\Delta E$ with an energy change $\Delta E$ from the initial to final times caused by external perturbation during $0\leq t\leq T_0$.
The validity of Eq.(\ref{diagonal}) depends on how accurately the conservation of energy $\langle E_m|U|E_n(0)\rangle\cong 0 (E_m\neq E_n(0)+\Delta E)$ holds.
The third line of Eq.(\ref{element}) behaves as a function of $E_n$ as 
\begin{eqnarray}
&&\langle E_n|U\rho(0)U^+|E_n\rangle \nonumber \\
&\cong&\frac{1}{\Omega(E(T))}\delta(E_n-E-\Delta E) \nonumber \\
&=&\langle E_n|\frac{1}{\Omega(E(T))}\delta(H(T)-E-\Delta E)|E_n\rangle \nonumber \\
&\cong&\langle E_n|\frac{1}{Z(T)}e^{-\hat{\beta}H(T)}|E_n\rangle, \label{diagonal} 
\end{eqnarray} 
where the density of the states at $t=T$ is determined uniquely from the normalization of Eq.(\ref{element}), $H(T)|E_n\rangle=E_n|E_n\rangle$ is used, and
 we replaced the microcanonical ensemble by the corresponding canonical ensemble whose density of the states has a sharp peak at $E+\Delta E$.
\end{itemize}
The relaxation property is also necessary for the thermodynamic quantities such as free energy difference considered in the nonequilibrium theorems to make sense\cite{Jarzynsky1,Jarzynsky2}.
\section{Entropy difference}
We estimate $\Delta S=D[U\rho(0)U^+||\rho({\it T})]$ based on the Eqs.(\ref{relaxation},\ref{diagonal}).
As a physical quantity $A$, we substitute $A=\log\rho({\it T})$.
Then Eq.(\ref{relaxation}) shows that 
\begin{equation}
{\rm Tr}U\rho(0)U^+\log \rho({\it T})\cong {\rm Tr}\rho({\rm T})\log\rho({\rm T}).
\end{equation}
And the entropy difference is calculated as 
\begin{eqnarray}
&&\Delta S\cong -{\rm Tr}\rho({\it T})\log\rho({\it T})+{\rm Tr}\rho(0)\log\rho(0) \nonumber \\
&=&{\rm Tr}\rho({\it T})H({\it T})-{\rm Tr}\rho(0)H(0)-\Delta F \nonumber \\
&=&\Delta S_{eq}. \label{entropy}
\end{eqnarray}
\section{Discussion}
In conclusion, by assuming that the initial state is described by canonical ensembles, we have derived an expression of the nonequilibrium entropy difference.   
We also revealed that the nonequilibrium entropy difference is well-approximated by the equilibrium entropy difference for typical state of macroscopic quantum systems.
In this way, the entropy increases for typical processes.  
\section{Acknowledgment}
The author is grateful to Professor A.Sugita for fruitful discussions.
This work is financially supported by JSPS research fellowship for young scientists under the Grant in aid 22$\cdot$7744.

\end{document}